\documentclass[12pt,preprint]{aastex}

\shorttitle{X-ray Flares from SMC X-1}
\shortauthors{Moon, Eikenberry, \& Wasserman}

\begin{document}

\title{SMC X-1 As An Intermediate-Stage Flaring X-ray Pulsar}

\author{Dae-Sik Moon, Stephen S. Eikenberry, \& Ira M. Wasserman}
\affil{Department of Astronomy \\
Space Science Building, Cornell University, Ithaca, NY 14853; \\
moon@astro.cornell.edu, eiken@astro.cornell.edu, ira@astro.cornell.edu}

\begin{abstract}
We present {\it Rossi X-ray Timing Explorer} observations of the X-ray pulsar SMC X-1.
The source is highly variable on short time scales ($<$ 1 h), 
exhibiting apparent X-ray flares occupying a significant fraction ($\sim$3 \%) of 
the total observing time, with a recurrence time of $\sim$100 s.
The flares seem to occur over all binary orbital phases, 
and correlate with the overall variability in the light curve.
We find a total of 323 discrete flares which have
a mean full width half maximum of $\sim$18 s. 
The detailed properties of SMC X-1 do not vary significantly 
between the flares and the normal state, 
suggesting that the flare may be an extension of the normal
state persistent emission with increased accretion rates. 
The flares resemble Type II X-ray bursts from GRO J1744--28.
We discuss the origin of the SMC X-1 flares in terms of 
a viscous instability near the inner edge of the accretion disk
around a weakly magnetized X-ray pulsar, and 
find this is consistent with the interpretation that SMC X-1 is
in fact an intermediate-stage source like GRO J1744--28.
\end{abstract}


\section{Introduction}

Neutron star X-ray binaries are generally
categorized into two groups: 
low mass X-ray binaries (LMXBs) and X-ray pulsars.
The surface magnetic field of the central neutron star
in an LMXB is thought to be $\sim$10$^8$ G.
The mass accretion with this magnetic field is most likely spherical, 
so that no significant inhomogeneity in the
X-ray emission over the neutron star surface is expected
-- i.e., no persistent coherent pulsations are observed.
The strong magnetic field, $\sim$10$^{12}$ G, of an X-ray pulsar,
on the other hand, can funnel the accretion
matter onto the magnetic pole, which makes the central neutron star appear as a pulsar.

Of particular interest are the so-called ``intermediate-stage sources''
speculated to lie between the LMXBs and X-ray pulsars,
including  ``the Rapid Burster" (MXB 1730--355), 
GRO J1744--28 (``the Bursting Pulsar"), and SAX J1808.4--3658  (``the accreting millisecond pulsar") 
(Lewin et al. 1976; Fishman et al. 1995; in 't Zand et al. 1998).
Both the Rapid Burster and GRO J1744--28 exhibit Type II X-ray bursts; however,
only the former shows Type I X-ray bursts while only the latter has apparent coherent pulsations
(Lewin et al. 1996).
SAX J1808.4--3658, on the other hand, shows both Type I bursts
and coherent pulsations (Wijnands \& van der Klis 1998; Chakarbarty \& Morgan 1998), 
but not Type II bursts.
The magnetic field strengths of these sources have been inferred to be $\sim$10$^{8-11}$ G,
between those of LMXBs and X-ray pulsars.

Another possible intermediate-stage source is the X-ray pulsar SMC X-1,
which has similar properties to GRO J1744--28, including 
its fast spin period ($\sim$0.72 s for SMC X-1; $\sim$0.47 s for GRO J1744--28),
steady spin-up, and inferred magnetic field ($\sim$10$^{11}$ G ) 
(Bildsten \& Brown 1997; Li \& van den Heuvel 1997).
In addition, once SMC X-1 was observed with an X-ray burst 
that resembles Type II bursts (Angelini, Stella, \& White 1991).
It may be possible, therefore, that SMC X-1 and GRO J1744--28
belong to a distinctive group of X-ray binaries, ``bursting pulsars",
which show both coherent pulsations and Type II X-ray bursts (Li \& van den Heuvel 1997).
To investigate this important possibility,
we analyze all publicly available RXTE data for SMC X-1,
searching for phenomena that may be related to X-ray bursts.  
We report that SMC X-1 in fact exhibits active flares resembling 
Type II bursts from GRO J1744--28.

\section{Data Analysis and a Flare Search}

We analyzed all publicly available RXTE Proportional Counter Array
observations toward SMC X-1.
Photon arrival times from the Good Xenon mode
were transformed to the solar system barycenter using the JPL DE400 ephemeris.
The Very Large Event Models were used to subtract backgrounds,
and only the Standard 2 data obtained from the top xenon
layers of Proportional Counter Units 0, 1, and 4 were considered for spectral analysis.
Both the data within 30 minutes after passages through South Atlantic Anomaly 
and/or with a high ($>$ 0.1) electron ratio were ignored.

The data toward SMC X-1 are occasionally contaminated
by outbursts of the nearby ($\sim$27$'$ away) 
transient source XTE J0111.2--7317 (Chakrabarty et al. 1998). 
The contamination was easily identified by the source's $\sim$31-s pulsations,
with powers reach up to $>$ 5000 in Leahy-normalized PDSs.
We searched for weak contamination in PDSs as narrow
statistically significant ($>$ 10 in Leahy-nomalized PDSs) peaks
at $\sim$0.032 Hz (and/or at its harmonic frequencies).
We excluded any data contaminated by XTE J0111.2--7317 from our analyses,
as well as data containing dip-like features (e.g., Her X-1; Moon \& Eikenberry 2001)
or close to the eclipse (i.e., binary orbital phase between 0.9 and 0.1) of SMC X-1.
We obtained a total of 150 data segments with an average length of $\sim$1360 s,
giving a net observational duration of $\sim$204 ks.
For spectral analysis, we considered only the spectrum between 2.5--25 keV range 
due to a PCA responsivity problem (R. Remillard 2001, private communication),
and assumed a systematic uncertainty of 1 \%.

In our analysis, we define ``flares'' to be the part of a light curve
that has three or more consecutive data bins with photon counts
larger than a threshold value of 3$\sigma$ Poisson noise
above the mean photon count in a given light curve.
We analyzed all the 150 light curves as follows searching for the flares.
First, we binned each light curve to 4-s time resolution and made a flare list.
We extended the search with 2- and 8-s time resolutions,
excluding the flares already found with different time resolutions.
We found a total of 323 flares,
and fit them a Gaussian function to estimate width and peak intensity.
We obtained $\sim$18 s and $\sim$15 s for the mean and rms standard deviation of FWHM; 
the reduced $\chi^2$ (= $\chi^2_{\nu}$) of the fits range from 0.5 to 1.9.
The total duration above half-maximum is $\sim$5.8 ks,
indicating that SMC X-1 spends $\sim$3 \% of its time on flares.

\section{Flare Examples and Correlations with Other Parameters}

The rms variability of the 150 light curves ranges between 6 and 16 \%,
with a mean of $\sim$11 \%.
Figure 1 presents three light curves (and their PDSs) 
with very low (7 \%), average (11 \%), and very high (16 \%) rms variability
as examples representing three different variability levels.
The mean photon count rates are 120.5 $\pm$ 8.5, 116.1 $\pm$ 12.9, and 146.4 $\pm$ 23.3, 
for Figure 1a, 1b, and 1c, respectively.
No flare is found in Figure 1a, which has a low rms variability; 
three flares are found Figure 1b, which has an average rms variability. 
Strong flaring activity is well illustrated in Figure 1c, 
which has a high rms variability, with several apparent flares lasting for a few tens of seconds.
At the flare peak, the photon count rate rises up to
$\sim$2.5 times of that outside the flares.
The inset shows a Gaussian fit to the intense
flare at $t$ $\simeq$ 900 s, with an FWHM estimated to be $\sim$23 s.
Leahy-normalized PDSs of the three light curves (Figure 1d, 1e, and 1f) 
show strong QPO-like peaks around 10 mHz, 
as well as the peaks caused by the source's coherent pulsations 
at $\nu$ $\simeq$ 1.4 Hz (and their harmonics),
indicating the existence of $\sim$100-s aperiodic variability
independent of the flaring activity.
The fractional rms (FRMS) amplitudes in the 2--50 mHz
range are 7 $\pm$ 1, 9 $\pm$ 1, and 14 $\pm$ 1 \% for 1d, 1e, and 1f, respectively.

Figure 2 presents the pulse profiles of the three light curves,
all showing the double-peaked, smooth profile typical of SMC X-1 (e.g., Levine et al. 1993).
They have the same pulsed fractions of $\sim$39 \%,
with their second peaks at phase 0.56 with respect to the first ones at phase 0.
The ratio of the second peak to the first peak
is $\sim$0.78 for Figure 2a, while it is $\sim$0.96 for 2b and 2c.
No significant variation in the pulse peak ratio has been found 
within a given light curve, 
regardless whether it is obtained inside or outside flares.

Figure 3 compares the phase-averaged softness ratio, 
defined to be the ratio of the soft X-ray (2--5 keV) photon count 
rates to those of the hard X-ray (5--13 keV),
of the three light curves in Figure 1
with the total photon count rates at 2--25 keV with 4-s resolution.
The average softness ratio of the three light curves is invariant:
0.47 $\pm$ 0.03, 0.47 $\pm$ 0.02 , and 0.46 $\pm$ 0.04 for Figure 1a, 1b,
and 1c, respectively. 
This contrasts with the softness ratio of the LMC X-4 flares
(Moon, Eikenberry, \& Wasserman 2002), 
which shows a strong linear correlation with the total photon count rate (inset in Figure 3).

We examined the spectral invariance of SMC X-1 over the flaring activity
indicated by the constant softness ratio distribution more thoroughly
via fitting the 32-s spectra obtained from the largest peak in each of the 
three light curves
(i.e., the peak at $t$ $\simeq$ 900 s in Figure 1a, at $t$ $\simeq$ 1050 s in 1b, 
and at $t$ $\simeq$ 1110 s in 1c) to a model spectrum.
The model spectrum consists of a power-law component with a high-energy cutoff 
(for non-thermal magnetospheric emission)
and a Gaussian component (for iron line emission),
together with a component for photoelectric absorption 
by intervening interstellar matter. 
We fixed the central energy of the Gaussian component to be 6.7 keV 
based on previous results \cite{aet91}.
Figure 4 compares the observed spectra with the best-fit model spectra, 
and Table 1 lists the best-fit parameters and $\chi^2_{\nu}$ of the fits.
Although all parameters are poorly constrained,
the power-law index does not change significantly over the flaring activity --
consistent with the softness ratio distribution (Figure 3).
The difference in $N_{\rm H}$ may be caused by the motion of the precessing, tilted accretion
disk suggested to be responsible for the super-orbital motion of SMC X-1,
but the low energy limit of the spectral fits (i.e., 2.5 keV) 
makes it difficult to constrain $N_{\rm H}$ propertly,
because the photoelectric absorption by intervening interstellar matter
is expected to be most significant in the soft energy band.

In order to perform statistical analyses, we calculated the flare fraction,
which we define to be the ratio of the integrated time that 
SMC X-1 is flaring (i.e., within FWHMs of the Gaussian fits)
to the total observing time of a given parameter,
and investigate its correlation with the parameter.
Figure 5 shows the distribution of the flare fraction as functions of
the binary orbital phase (5a), rms variability of the light curve (5b),
and the pulse peak ratio (5c).
Some important results are worth noticing: the flare fraction
(1) is larger than 1.4 \% over all the orbital phases 
(with its minimum at the orbital phase of $\sim$0.35),
(2) shows a significant variation from phase to phase, and
(3) increases with the rms variability of the light curve, 
as well as with the pulse peak ratio.
The correlation between flare fraction and rms variability 
identified in Figure 5b remains very similar
when we use the flare-subtracted rms variability.

\section{Discussion} 

SMC X-1 shows active flares that occupy $\sim$3 \% of the total observing time. 
The flares seem to occur over all binary orbital phases,
and the flaring activity is proportional to the rms variability of the light curve (Figure 5). 
Except for the small change in the pulse peak ratio (Figure 2 and 5),
no significant change is found along with the flaring activity.
All these suggest that the SMC X-1 flares may be
simple extensions of the persistent emission of a normal state with increased accretion rates,
but without significant changes in the geometry of the accretion flows and the magnetosphere.
Under the hypothesis that the double peaks in the pulse profile (Figure 2)
are due to the two magnetic poles of SMC X-1,
one simple explanation for the change in the pulse peak ratio may be that
the increase in the accretion rate onto the fainter pole 
is higher than that onto the brighter pole during the flares.
The invariant pulse peak ratio in a given light curve (regardless of flares),
together with the correlation between the flare-subtracted rms variability
and the flare fraction, on the other hand, may indicate the existence
of variability whose time scale is longer than the flare recurrence time, $\sim$100 s.

The SMC X-1 flares differ from Type I X-ray bursts from LMXBs for various reasons,
including the shape of profiles and spectral properties.
While the profile of Type I bursts shows an abrupt increase with 
an exponential decay in most cases, 
the SMC X-1 flares have symmetric Gaussian shapes.
The X-ray spectrum of the SMC X-1 flare is far from thermal spectrum of Type I bursts
(although the thermal spectrum of an X-ray pulsar is not well constrained, 
so it is not completely excluded that the SMC X-1 flare spectrum is thermal).
In addition, the SMC X-1 spectrum does not show any apparent variation within a flare,
while Type I burst shows spectral cooling as the burst continues.
The changes in the softness ratio and pulse profile of the SMC X-1 flares
differ from those of the LMC X-4 flares (Figure 3; Moon et al. 2002),
which is one of the most well known and regular flaring X-ray binaries.

The SMC X-1 flares recall the Type II X-ray bursts found in GRO J1744--28 
mainly due to the spectral invarince over the flaring activity.
In fact, the burst spectrum of GRO J1744--28 was found to be very similar to 
that of SMC X-1, with a similar photon index of $\sim$1.2 and 
e-folding energy of $\sim$14 keV (Lewin et al. 1996).
One difference is that the SMC X-1 flares lack the post-flare dip 
which often follows the Type II bursts from GRO J1744--28.
The rather gradual rise of the SMC X-1 flares, however, 
may be responsible for it via offering sufficient time 
to replenish the material in the accretion disk. 
This is consistent with the interpretation that the SMC X-1 flares
are simple extensions of a normal state with increased accretion rates,
which is probably the most compelling argument for attributing the origin of 
the SMC X-1 flares to an accretion disk instability.
Given the difference between GRO J1744--28 and SMC X-1
(i.e., a transient low-mass system versus a persistent high-mass one),
we consider that the magnetic field strength, suggested to be
comparable for the two sources, is critical to understanding
the Type II bursts from GRO J1744--28 and the X-ray flares from SMC X-1.

We note that SMC X-1 may be capable of experiencing
a viscous instability, namely the Lightman-Eardley instability (Lightman \& Eardley 1974),
due to its $\sim$10$^{29}$ G cm$^3$ magnetic moment \cite{lv97}. 
The radiation pressure in this case is comparable to the
gas pressure around the inner disk radius (i.e., the magnetospheric radius),
resulting in a viscous instability with slightly 
increased mass accretion (e.g., Cannizzo 1996, 1997).
Because the instability develops near the inner edge of the accretion disk,
it has an advantage in explaining bursts/flares with a short recurrence time.
The viscosity parameter ($\alpha$) of the classical $\alpha$-disk (Shakura \& Sunyaev 1973)
at the transition radius between the ``inner region" and the ``middle region" 
around an 1.4 $M_{\odot}$ neutron star is
$\alpha \simeq 216 \; t_{\rm vis}^{-3/2} \dot M_{17}$,
where $t_{\rm vis}$ is the viscous time scale in seconds and  $\dot M_{17}$ is 
the mass accretion rate in units of 10$^{17}$ g s$^{-1}$.
For a typical value $\dot M_{17}$ $\simeq$ 20 for SMC X-1 (e.g., Wojdowski et al. 2000),
the viscosity parameter is $\alpha$ $\simeq$ 0.14 and 4.4 when 
$t_{\rm vis}$ is 1000 and 100 s, respectively.
The value 4.4 for $t_{\rm vis}$ = 100 s is somewhat larger than generally expected, 
$\alpha$ $\lesssim$ 1.
However, even larger values ($\alpha$ $\simeq$ 10--100) were obtained for the dwarf nova HT Cas, 
possibly due to the patch nature of the accretion disk (Vrielmann, Hessman, \& Horne 2002).
Alternately, the strong magnetic field of SMC X-1 may help increase the viscosity parameter
(R. Lovelace 2002, private communication).
We need detailed numerical studies to investigate this scenario more thoroughly.

\section{Summary and Conclusions} 

Through the analysis of the all publicly available RXTE data toward the
X-ray pulsar SMC X-1, we find that the source is highly variable on short time scales ($<$ 1 h), 
and that Gaussian flares occur over all orbital phases with a $\sim$100 s recurrence time scale.
The flares occupy $\sim$3 \% of the total observing time, and the flaring activity 
is proportional to the overall variability of the source.
While the pulse peak ratio shows a small systematic change along with the flaring activity,
the PDSs, pulse profiles, softness ratios, and X-ray spectra
during the flares are very similar to those outside the flares,
indicating that the flares are probably extensions of a normal state just with
increased accretion rates. 
This supports the interpretation that the SMC X-1 flares have their origin in an accretion 
disk instability and the suggestion that it may belong to a distinctive group
of ``bursting pulsars" with the ``bursting pulsar" GRO J1744--28,
owing to its $\sim$10$^{11}$ G surface magnetic field. 
A viscous instability near the inner-edge of the accretion disk
might be responsible for the SMC X-1 flares, although detailed studies on this scenario 
need to be done in the future.

\acknowledgments
We would like to thank the anonymous referee for the useful comments and suggestions.
D.-S. M. acknowledges Akiko Shirakawa, Dong Lai, and Richard Lovelace for their comments. 
This research has made use of data obtained from the {\it
High Energy Astrophysics Science Archive Research Center}
provided by NASA's Goddard Space Flight Center.
D.-S. M. is supported by NSF grant AST-9986898.
S. S. E. is supported in part by an NSF Faculty Early Careeer Development
award (NSF-9983830). 
I. M. W. is supported by the NASA grant NAG-5-8356.

\clearpage
\begin{deluxetable}{lrrr}
\tablecolumns{4}
\tablewidth{0pt}
\tablecaption{Best-fit Parameters for Spectra Shown in Figure 4 \label{tbl-1}}
\tablehead{
\colhead{Parameter} & \colhead{(a)}     & \colhead{(b)}  & \colhead{(c)}}
\startdata
$N_{\rm H}$\tablenotemark{*} (10$^{22}$ cm$^{-2}$)                     & 3.2(0.9)  & 1.6(1.5)   & 2.2(0.9)  \\
$\alpha$\tablenotemark{*}                                  & 1.6(0.3)  & 1.3(0.5)   & 1.6(0.3)  \\
$E_{\rm c}$ (keV)\tablenotemark{*}                   & 17.2(7.1) & 6.9(3.2)   & 14.3(5.2) \\
$E_{\rm f}$ (keV)\tablenotemark{*}                   & 7.8(6.9)  & 12.6(10.2) & 14.0(10.6)  \\
$\chi^2_{\nu}$\tablenotemark{**}                       & 0.81      & 1.3        & 0.81      \\
Flux (10$^{-9}$ ergs cm$^{-2}$ s$^{-1}$)              & 1.7       & 2.0        & 2.8       \\
\enddata
\tablenotetext{*}{$N_{\rm H}$ and $\alpha$ are the hydrogen nuclei column density 
of the intervening matter and the index of the power law component.
$E_{\rm c}$ and $E_{\rm f}$ represent the cutoff and e-folding energy
of the high-energy cutoff component.}
\tablenotetext{**}{Each spectrum has 48 degrees of freedom.}
\tablecomments{Energy range is 2.5--25 keV, and the 90 \% uncertainty levels are quoted in the parentheses.}
\end{deluxetable}

\clearpage
\begin{figure}
\plotone{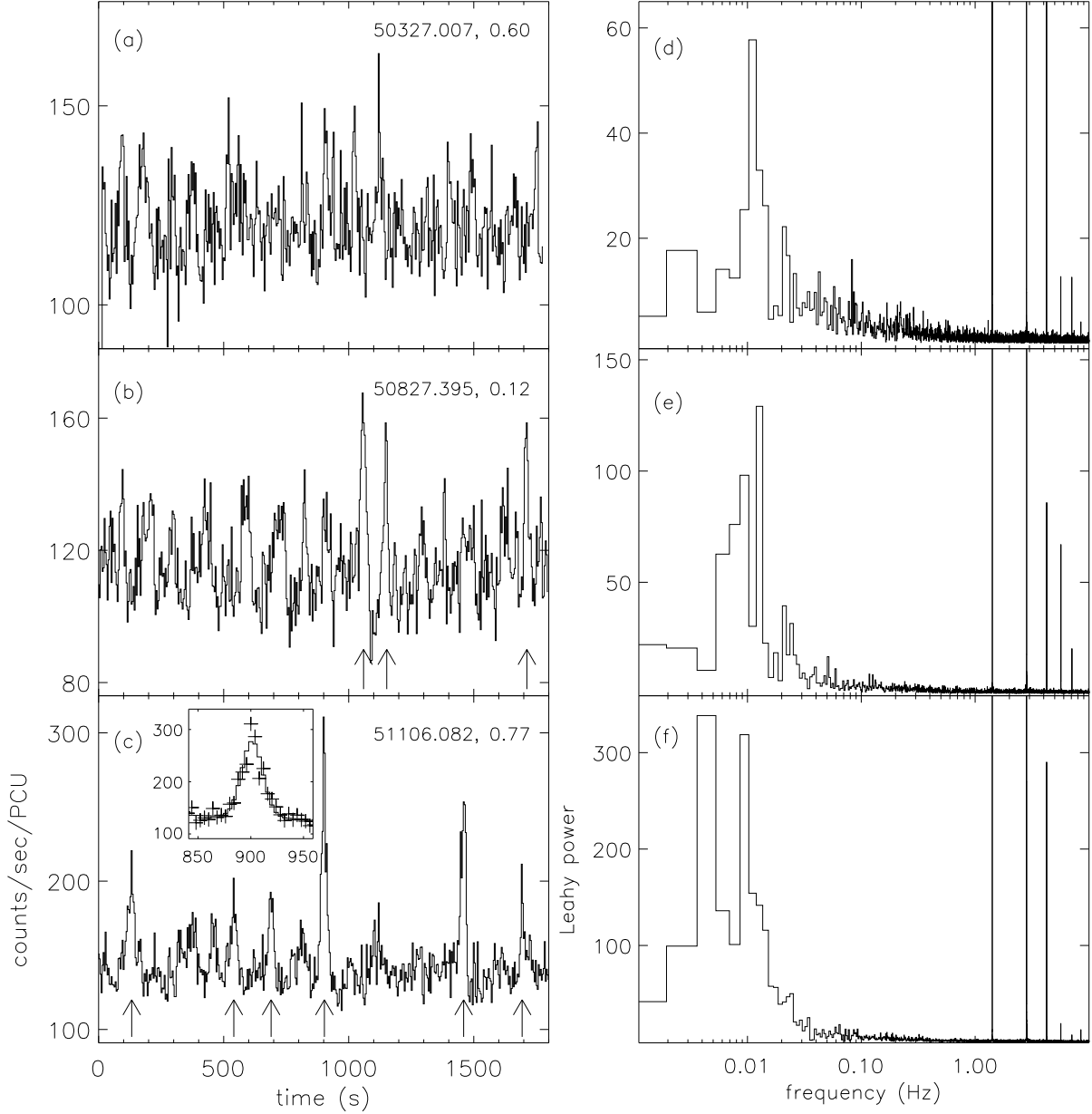}
\caption{
(a--c): 
Light curves (in the 2--25 keV range) with 4-s time resolution 
of three data segments that have $\sim$7 \% (a), $\sim$11 \% (b), and
$\sim$16 \% (c) rms variabilities.
The numbers in the upper-right corners represent the
MJD of the beginning of each light curve and its binary orbital phase.
The inset in (a) compares the intense flare (crosses) at $t$ $\simeq$ 900 s with
a Gaussian fit (solid histogram). The arrows indicate the central positions of nine flares. 
(d--f): 
Leahy-normalized PDSs of the three light curves in the left-hand side.
\label{fig1}}
\end{figure}

\clearpage
\begin{figure}
\plotone{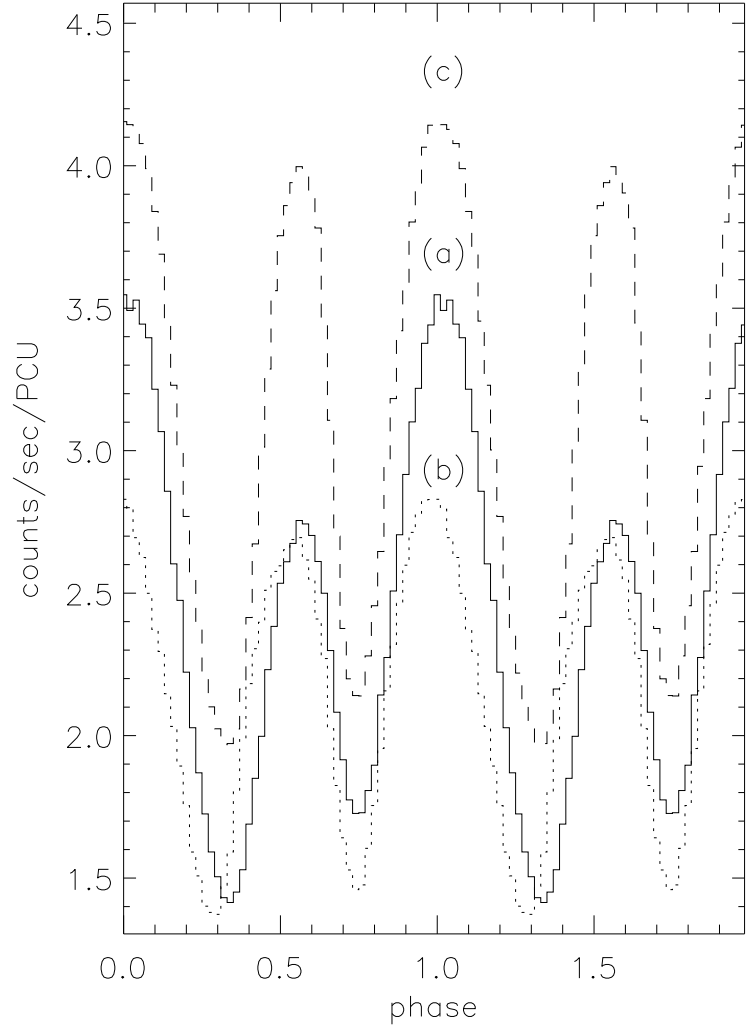}
\caption{Pulse profiles of the three light curves in Figure 2
-- solid line for 2a, dotted line for 2b, and dashed line for 2c.
The number of bin is 50.
\label{fig2}}
\end{figure}

\clearpage
\begin{figure}
\plotone{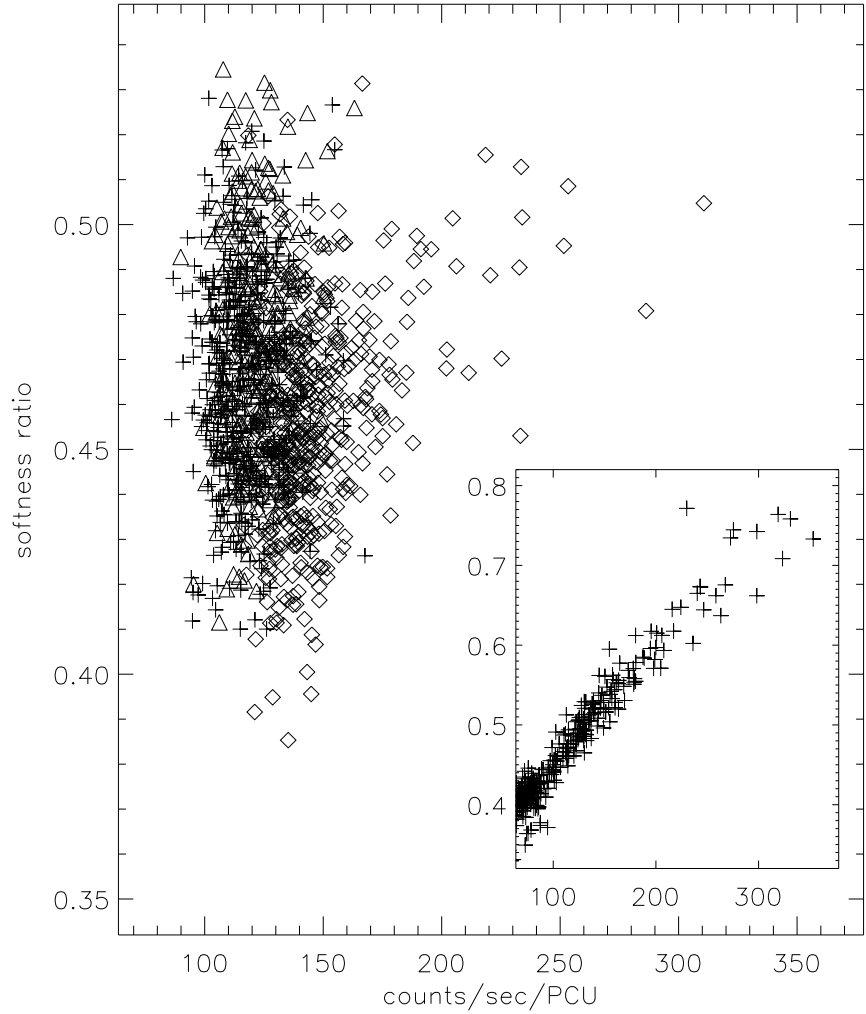}
\caption{Softness ratio of the three light curves in Figure 1: triangle for
1a, crosses for 1b, and diamonds for 1c. The softness ratio is a ratio
of the soft X-ray (2--5 keV) photon count rates to those of hard X-ray (5--13 keV).
The inset shows the same softness ratio of the LMC X-4 flares (Moon et al. 2002).
All the softness ratios are obtained with 4-s resolution.
\label{fig3}}
\end{figure}

\clearpage
\begin{figure}
\plotone{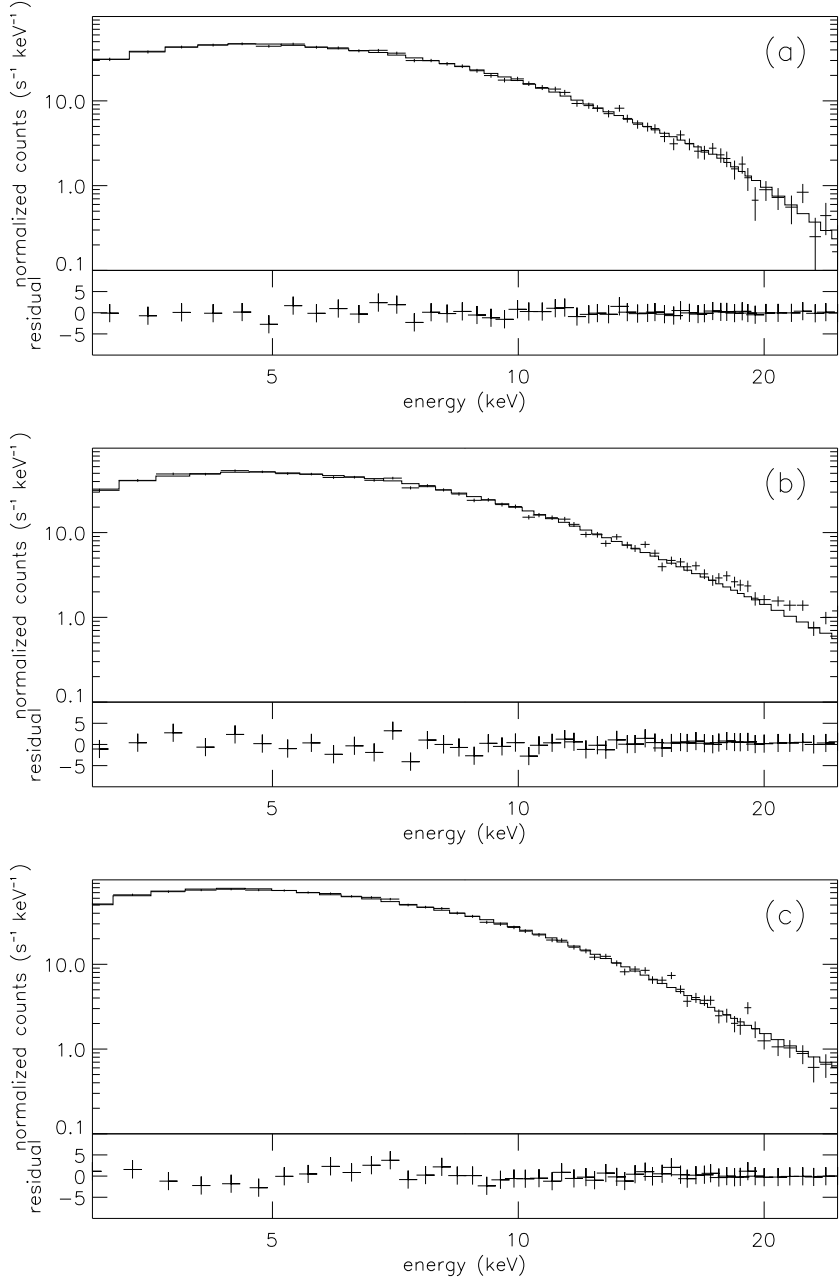}
\caption{Comparison of the observed spectra (crosses) of the brightest peaks
of the three light curves in Figure 1
(the peak at $t$ $\simeq$ 900 s in 1a, $t$ $\simeq$ 1050 s in 1b, and $t$ $\simeq$ 1110 s in 1c)
with the spectra (solid histogram) of the best-fit parameters.
The bottom panels represent the residuals of the fittings.
\label{fig4}}
\end{figure}

\clearpage
\begin{figure}
\plotone{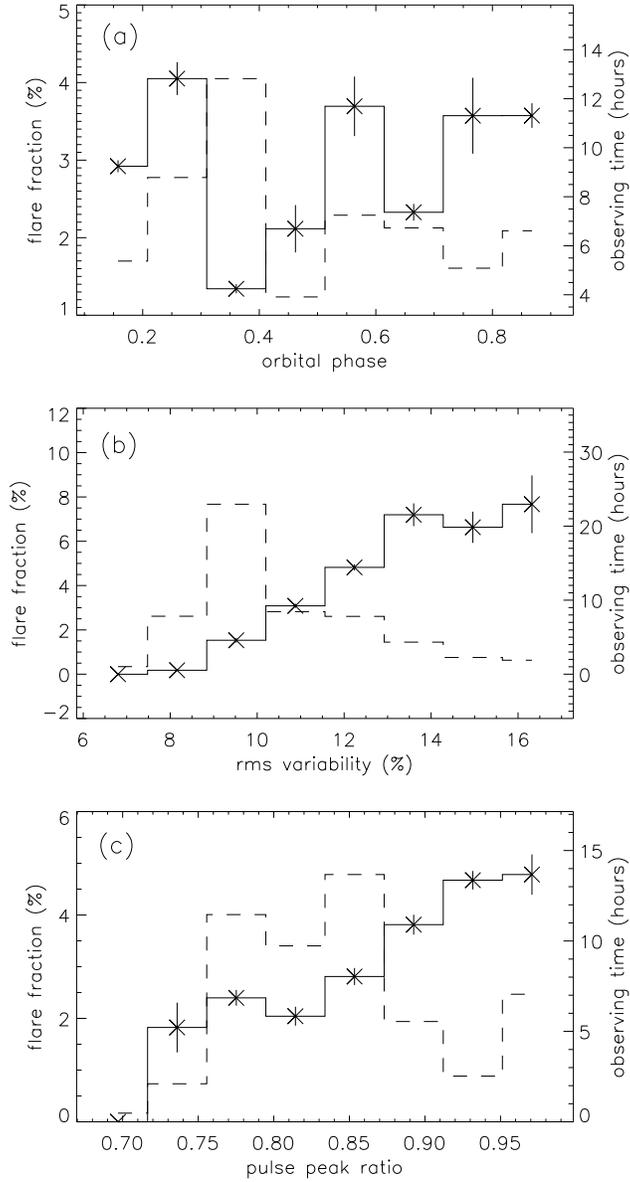}
\caption{The distributions of the flare fraction (solid histograms) versus
the binary orbital phase (a), the rms variability of the light curve (b),
and the pulse peak ratio (c).
The error bars correspond to 1-$\sigma$ confidence level.
The dashed histograms represent the integrated observing time (the right-hand ordinates)
in the corresponding binary orbital phase (a), the rms variability (b), and the pulse peak ratio (c).
\label{fig5}}
\end{figure}

\end{document}